\documentclass[fleqn,10pt]{wlscirep}
\usepackage[utf8]{inputenc}
\usepackage[T1]{fontenc}
\usepackage{multicol}
\usepackage{float}

\graphicspath{{./Figures/}}

\usepackage{xcolor}

\title{Observation of open scattering channels}

\author[1]{Reinier van der Meer}
\author[2]{Michiel de Goede}
\author[2]{Ben Kassenberg}
\author[2]{Pim Venderbosch}
\author[2]{Henk Snijders}
\author[2]{J\"{o}rn Epping}
\author[2]{Caterina Taballione}
\author[2]{Hans van den Vlekkert}
\author[1,2]{Jelmer J. Renema}
\author[1,*]{Pepijn W.H. Pinkse}

\affil[1]{MESA+ Institute, University of Twente, PO Box 217, 7500 AE Enschede, The Netherlands}
\affil[2]{QuiX Quantum BV, Enschede, The Netherlands}

\affil[*]{p.w.h.pinkse@utwente.nl}

\keywords{DMPK, Scattering, Integrated optics, Random matrix theory}

\begin{abstract}
The existence of fully transmissive eigenchannels ('open channels') in a random scattering medium is a counterintuitive and unresolved prediction of random matrix theory. 
The smoking gun of such open channels, namely a bimodal distribution of the transmission efficiencies of the scattering channels, has so far eluded experimental observation. 
We observe an experimental distribution of transmission efficiencies that obeys the predicted bimodal Dorokhov-Mello-Pereyra-Kumar distribution. 
Thereby we show the existence of open channels in a linear optical scattering system.
The characterization of the scattering system is carried out by a quantum-optical readout method. 
We find that missing a single channel in the measurement already prevents detection of the open channels, illustrating why their observation has proven so elusive until now. 
Our work confirms a long-standing prediction of random matrix theory underlying wave transport through disordered systems.
\end{abstract}
\begin{document}

\flushbottom
\maketitle
\thispagestyle{empty}
\begin{multicols}{2}

\noindent Wave transport through scattering media is ubiquitous in nature and technology. Its physics is essential in electron transport in quantum dots and nanowires\cite{Lee1985,Murphy2007}, conductance fluctuations in electron transport\cite{Washburn1988}, optical transmission in multimode fibers\cite{xiong_2018_LightSciAppl}, the theory of acoustic waves\cite{Sprik2008,Aubry2009,Gerardin2014}, and fluctuations in light transport\cite{deBoer1994,Popoff2010,Kim2012}. An understanding of scattering physics, together with adaptive optical technologies, allows us to exploit, scattering for various applications such as wavefront shaping\cite{Vellekoop2007}, physical unclonable functions\cite{Pappu2002,Goorden2014,Uppu2019}, communication\cite{Skipetrov2003}, and imaging\cite{Vellekoop2010}. 

\begin{figure*}
    \begin{center}
    \includegraphics[width=1\linewidth,keepaspectratio]{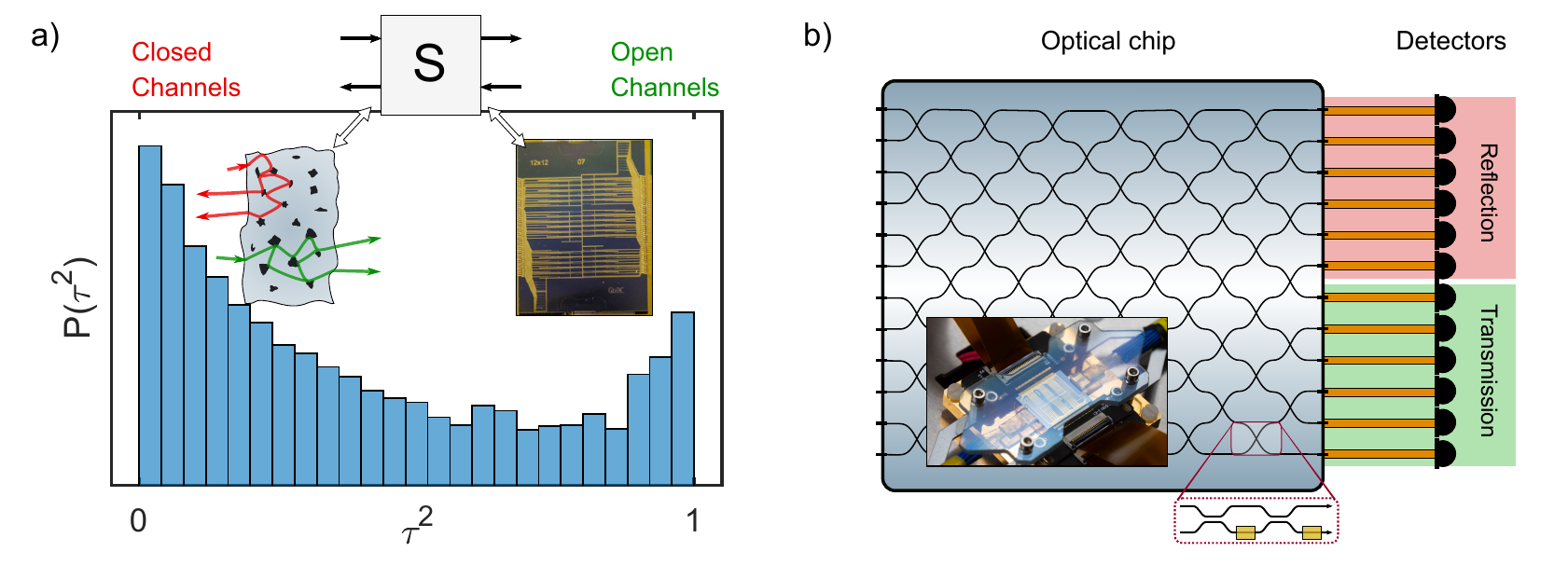}
    \caption{\textbf{Mapping a scattering system with open channels.}a) The bimodal distribution of the singular values $\tau$ of the transmission matrix of a loss-free scattering system is described by the DMPK distribution. The peak near $\tau^2=0$ (left, on red background) is caused by the closed channels and the peak near $\tau^2=1$ (right, green background) is caused by the open channels. The inset shows how light incident on a scattering medium tends to be fully reflected (red), but some eigenmodes have a near-unity transmission (green). This system can be completely modeled by a scattering matrix $\mathbf{S}$ that we simulate on a programmable optical network.
    b) A schematic of the programmable optical network with $12$ input and output modes that implements the scattering matrix $S$. The top 6 output modes (red) correspond to the reflection modes of $S$ and the bottom (green) ones correspond to the transmission modes. On the bottom, a unit-cell of the network is shown. Each cell consists of two 50:50 splitters and two thermo-optical elements for tunability. }
    \label{fig:fig1Fancy}
    \end{center}
\end{figure*}

Open channels have a pivotal role in transport through disordered systems\cite{Rotter2017}. These open channels are eigenmodes of the transmission matrix with full transmission through an otherwise opaque medium\cite{Dorokhov1982,Mello1988}. It is not just the case that open channels can exist - one can easily imagine encountering a fully transmissive mode with an exponentially small probability -, but rather that transport is dominated by fully closed and fully open channels. This is because the distribution of transmission eigenvalues (more precisely, that of the singular-values) is {\it bimodal}, with one peak at low transmission values and one peak at high values, as indicated in Fig. \ref{fig:fig1Fancy}a). This bimodal distribution is considered to be one of the most spectacular predictions of random matrix theory\cite{Rotter2017}, and remains as yet unconfirmed in direct experiments.

Technologically, open channels underlie many of the applications of scattering systems. For example, it is possible to increase the transmission to (near) unity in a disordered medium by coupling the input light into one of the open channels. This allows for lossless transmission. One can take this one step further by using a spatial light modulator to 'undo' the scattering and create a focus behind the scattering sample\cite{Vellekoop2007}. As the transmission matrix is not unitary, simply applying some unitary matrix with the spatial light modulator does not guarantee a high transmission; a high transmission is only possible when open channels exist\cite{Goetschy2013,Popoff2014}. Furthermore, conductance fluctuations in optical or electronic transport intricately depend on the existence of the bimodal distribution and its higher-order moments \cite{Washburn1988,deBoer1994}.

Despite the central role of open channels in transport through disordered media, only indirect signatures for the existence of open channels have been provided\cite{Choi2011,Vellekoop2008,Shi2012,Gerardin2014}. The limiting factor in measuring the bimodal singular values distribution in scattering media is the difficulty to individually probe and measure all of the modes of the system\cite{Yu2013,Goetschy2013,Popoff2014}. This can be understood by realizing that long-range mesoscopic correlations at the output build-up as the light gets randomly scattered and interferes with itself. Missing modes imply losing this long-range order, which ultimately leads to uncorrelated Marcenko-Pastur (MP) statistics. The open channels are only observable when the number of controlled modes must be $>\approx95\%$ of the total number of channels\cite{Popoff2014,Marcenko1967}. Despite considerable effort, experimental access to a sufficient fraction of modes has so far not been achieved. Reimposing unitarity on only the observed modes amounts to the assertion that the observed set of modes is decoupled from all others, which is unjustified in the experimental situation of a scattering system. Consequently, the second peak in the singular-value distribution has not unambiguously been observed yet.

In this work, we report experimental proof of the existence of open channels from a telltale high-transmission peak in the singular-value distribution (SVD). We do so by mapping a scattering medium with exactly 6 input and 6 output channels to a $12\times12$ scattering matrix implemented on a linear integrated optical processor. We experimentally characterise the full transmission matrix using two-photon interference as a robust readout technique. From this, we observe the bimodal transmission singular value distribution. Profiting from the superb access and control over all modes given by an integrated photonic processor, we observe that open channels are only visible in the experimental eigenvalue distribution when all modes are considered.

The natural mathematical framework describing this scattering physics is random matrix theory (RMT). RMT replaces system-specific details with a scattering matrix
 $\mathbf{S}$
\begin{equation}
    \mathbf{S} = \begin{bmatrix}
    R &T\\ T' & R'\\
    \end{bmatrix},
\end{equation}
where the submatrices $T$ and $R$ are the transmission and reflection matrices, respectively. This scattering matrix contains the appropriate statistical properties of the system, while remaining agnostic to the microscopic details of the scatterer. This allows to study their physics on any system that captures these statistics.

\begin{figure*}[hbt!]
\includegraphics[width=1\textwidth,keepaspectratio]{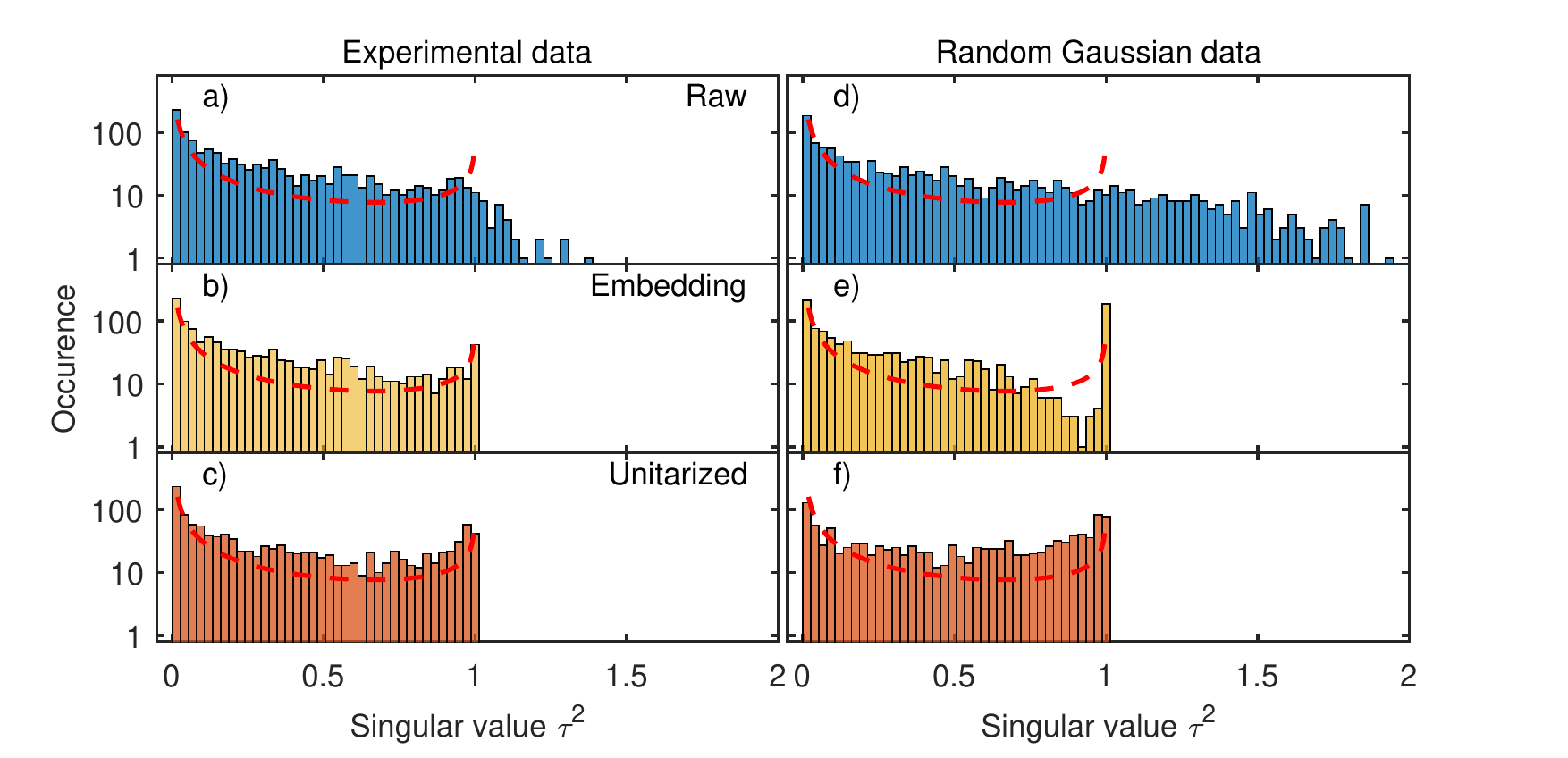}%
\caption{\textbf{Open channels.} The singular values distributions of the transmission matrix for two data sets and three different methods of processing. The left three panels correspond with the actual experimental data and the right column with data from random Gaussian matrices. The top row (a and e) shows the raw data. The middle panels (b and e) show that the bimodal distribution can be recovered by embedding the matrices, but only for the experimental data. The bottom panels (c and f) show that the unitarization of the data always results in a bimodal distribution. The unitarized Gaussian matrices lead to a symmetric distribution, whereas the unitarized experimental data show the asymmetric DMPK distribution.}
\label{fig:fig3DataRaw}
\end{figure*}

We simulate our diffusive system on such a state-of-the-art $12$-mode one-way integrated photonic processor, shown in the inset of Fig. \ref{fig:fig1Fancy}a) and schematically in Fig. \ref{fig:fig1Fancy}b)\cite{Taballione2021}. On this network, an entire scattering matrix $\mathbf{S}$ is implemented, where the first six output modes are treated as 'reflection' modes and output modes $7-12$ as transmission modes.

Characterizing such a matrix only results in six singular values, which is not sufficient to build up the entire bimodal distribution. However, a major advantage of this network is that it is fully reconfigurable. For this experiment, we implemented a total of $200$ scattering matrices. The scattering matrices are generated by a numerical simulation of a $12$-mode scattering system with appropriate settings. The simulation is based on the method of\cite{Ko1988}, as described in more detail in the Supplemental materials.

Characterisation of the matrices on the network is performed by sending pairs of single photons into the network and sampling their output distribution with a battery of superconducting nanowire single-photon detectors (SNSPDs). Although it would in principle be possible to characterise the matrix with classical coherent light in an interferometrically stable setup, doing the readout with single photons has the advantage that we do not need interferometric stability of the fibers connecting the PIC network with the outside world\cite{Laing2012,Dhand2016}, a fact which arises from the phase-insensitivity of the single-photon quantum state. Hence our readout method is motivated by the quantum readout being more practical than the equivalent classical method.
 
The matrix amplitudes are sampled by sequentially injecting single photons into each input mode and measuring the output distribution. The photon flux is corrected for known experimental fluctuations such as the variations of pump power over time, relative detector efficiencies, and output losses of the chip. The phases of the matrix elements are characterized by sequentially measuring two-photon interference in the network for a given set of combinations of two input and two output modes\cite{Laing2012}.. 

To reduce experimental measurement time, we only characterized the phases of the transmission matrix, not of the reflection matrix. The matrix amplitudes are measured for the entire $\mathbf{S}$ matrix so that the $1$-photon output distribution can be normalised. See Methods for the details on the chip and setup.

\paragraph*{Open Channels.}

\begin{figure*}
    \centering
    \includegraphics[width=1\textwidth,keepaspectratio]{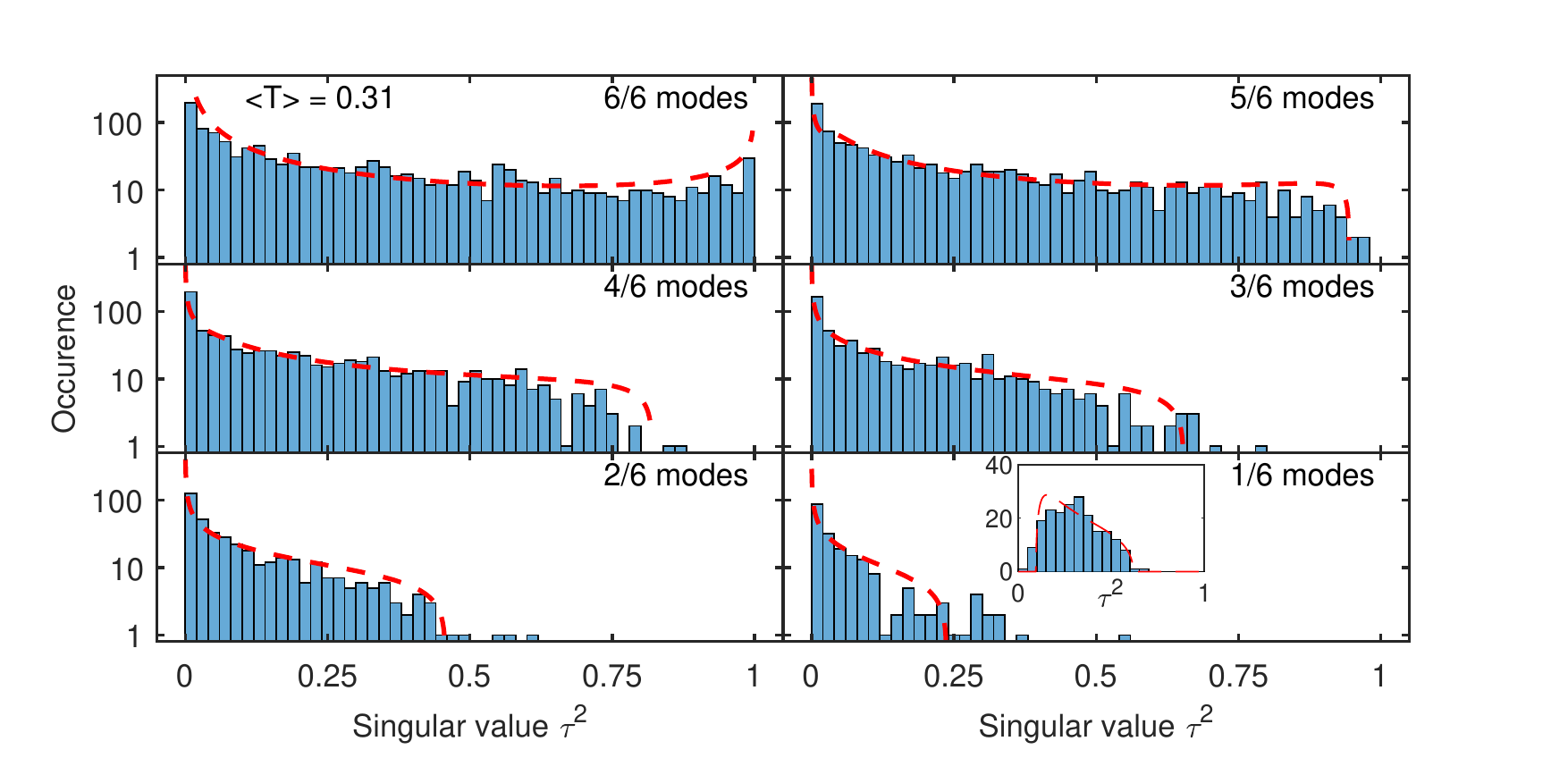}%
    \caption{\textbf{Mode Filtering.} The singular value distribution of the transmission matrix for the embedded matrices. Each panel corresponds with a different number of 'observed' modes. The observed distributions match the (zero-free-parameter) predictions from Goetschy and Stone\cite{Goetschy2013} well, as indicated by the red dashed lines. The insert in the last panel shows the observation and theory predictions for the situation where the control over the input channels is not reduced, a situation in which the (non-square) transmission matrix starts asymptotically resembling a random matrix.}
    \label{fig:fig4ModeFilter}
\end{figure*}

Figure \ref{fig:fig3DataRaw}a) shows the experimentally obtained singular-value distribution of the transmission matrix. The shoulder near the singular value $\tau^2=1$ in Fig. \ref{fig:fig3DataRaw}a) is indicative of the expected peak, this by itself is not enough to claim observation of open channels. Ideally, energy conservation results in singular values between $0$ and $1$. However, experimental noise resulted in a non-normalised $S$ matrix. The intensities of the rows sum up to $1.00 \pm 0.053$, whereas the columns sum up to exactly $1$ due to the normalisation of the measured output distribution.

The open channels are recovered in Fig. \ref{fig:fig3DataRaw}b) using a no-gain assumption by embedding the transmission matrix in a larger, unitary matrix. This embedding matrix can physically be understood as a matrix that also incorporates the losses and coupling to the environment\cite{Brod2020}. We can then apply the no-gain assumption to this larger embedding matrix and again extract the new transmission matrix $T$. More information on the embedding method can be found in the Supplemental Materials. 

The singular value distribution now has a large and relatively broad peak at $\tau^2=0$ and a smaller peak at $\tau^2=1$ indicating the open channels. Furthermore, the distribution follows the ideal DMPK curve indicated by the red, dashed line\cite{Beenakker1997}
\begin{equation}
    \rho(T) = A\frac{g}{2T\sqrt{1-T}},
    \label{eqDMPK}
\end{equation}
where $A$ is a zero-free-parameter scaling factor that converts the probability density function to counts. The extracted average transmission $\langle T \rangle = 31\% = \frac{l^*}{L}$ is close to the expected transmission ($37\%$). The value of the average transmission indicates that the system is approximately in the diffusive regime as $l^*<L$, where $l^*$ is the transport mean free path\cite{vanRossum1999} and $L$ the characteristic system size. The average dimensionless conductance $g = \sum_i \tau^2_i = 1.9 \pm 0.5$, where the uncertainty indicates the standard deviation over all $200$ independent conductance values.

Another data processing option is to impose unitarity, or energy conservation on the experimentally reconstructed scattering matrices. This is shown in Fig. \ref{fig:fig3DataRaw}c). This panel shows that the bimodal behaviour with the open channels is again recovered.

It is tempting to simply unitarize the experimental scattering matrix to mitigate the noise. However, because the essence of observing open channels is to not miss any modes, applying unitarization to a noisy scattering matrix amounts to imposing the desired solution of the data as it artificially imposes (long-range) correlations in the transmission matrix. However, these new correlations do no longer correspond with a DMPK system, but with that of a chaotic cavity\cite{Beenakker2009}. To emphasise this effect, we will compare our data analysis procedure of our data with that of artificial data, generated by computer from random complex Gaussian matrices with the same mean and variance as observed in our experimental data. This artificial data is shown in Fig. \ref{fig:fig3DataRaw}d).

Figure \ref{fig:fig4ModeFilter}e) shows the singular value distribution of the random Gaussian data after applying the same embedding procedure of Fig. \ref{fig:fig3DataRaw}b). The high peak at $\tau^2=1$ almost reaches $200$ and is the result of the renormalization of almost all $200$ Gaussian matrices. Despite the presence of the high peak at $\tau^2=1$, the distribution still clearly does not follow the one expected from DMPK statistics, which proves that it is possible to differentiate between actual and random data when the embedding procedure is used. This highlights the robustness of our data processing.

Finally, Fig. \ref{fig:fig3DataRaw}f) confirms the insight that the unitarization of random Gaussian matrices indeed results in a bimodal distribution of the singular values. The resulting distribution has lost its asymmetry, but this subtlety is eluded in experiments with limited data to sample the distribution.

\paragraph{Mode filtering.}
The observation of open channels in Fig. \ref{fig:fig3DataRaw} is only possible because of the complete control over the number of modes. Missing out even one mode is already sufficient to hide the open channels\cite{Goetschy2013}. Figure \ref{fig:fig4ModeFilter}b-f) shows that the correlations inside the transmission matrix disappear when the fraction of observed modes at both the input and the output is decreased. The resulting filtered distributions match the predictions of Goetschy and Stone\cite{Goetschy2013}, which are indicated by the red dashed lines. This emphasizes the demanding restriction that almost all modes must be included in measurement in order to observe the open channels.

When the fraction of either controlled input or output modes is decreased, then the singular values will become uncorrelated and the open channels again disappear. The singular-value distribution will asymptotically follow the Marcenko-Pastur (MP) law since this describes the singular values of random rectangular Gaussian matrices\cite{Marcenko1967}. The inset in the bottom right panel of Fig. \ref{fig:fig4ModeFilter} shows the observed distribution associated with the $1\times 6$ rectangular matrices. The Goetschy-Stone prediction is drawn in red for reference. 
The distribution already shows a maximum at intermediate singular values, a key characteristic of the MP distribution.

\paragraph{Discusssion}
In summary, we have successfully solved a long-standing problem by showing experimental proof of the bimodal behaviour of the transmission singular values of scattering systems. This was enabled by having access to all input and output modes of our system. The singular value distributions with a reduced number of modes follow the predictions of Goetschy and Stone \cite{Goetschy2013} with a near-perfect quantitative agreement. This confirms the long-standing hypothesis that the open channels can only be recovered when all modes can experimentally be accessed.
Our work fits in a trend of using well-defined photonic systems to investigate scattering physics \cite{Gilead2015,Crespi2013,harris_2017_Nat.Photonics,Stuetzer2018}. Our large, low loss and fully tunable processor can be used to study, for example, multi-photon interference effects in disordered systems, universal conductance fluctuations, or Anderson localisation \cite{Rotter2017}.
Another future direction could be to use recirculating mesh design, which allows for a more natural correspondence to the physical scattering systems.

\newpage
\section{Methods}
    \label{sec:methods}
    
\begin{figure*}
    \begin{center}
    \includegraphics{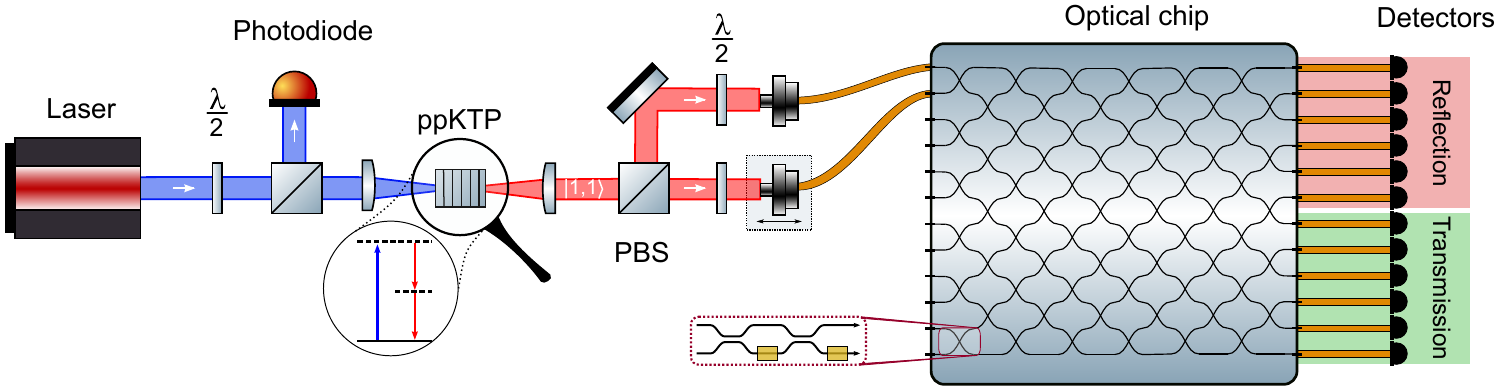}%
    \caption{\textbf{Setup.} A sketch of the setup. A pulsed laser is used to generate pairs of photons in a ppKTP crystal. The photons have orthogonal polarization and are separated by a polarizing beam splitter and subsequently coupled into a PM fiber which is connected to the optical network. After the optical network, the photons go through an SMF to the single-photon detectors via a fiber polarization controller (not shown). To guarantee temporal overlap of the photons, one of the fiber couplers is placed on a linear stage. A beam sampler is used to monitor the power using a calibrated photodiode and the pump beam is filtered out after the ppKTP crystal (not shown).}
    \label{fig:fig2Setup}
    \end{center}
\end{figure*}

The setup, shown in Fig. \ref{fig:fig2Setup}, generates pairs of photons in a Type-II degenerate spontaneous parametric downconversion (SPDC) source with a $2\,$mm periodically poled potassium titanyl phosphate (ppKTP) crystal. This crystal is pumped by a femtosecond mode-locked Ti:Sapphire laser (Tsunami, Spectra Physics) which emits light at $775\,$nm and has a linewidth of $5.5\,$nm. The generated photons are separated by a polarizing beam splitter and then injected into a polarization-maintaining fiber which routes these photons to the $\textrm{Si}_3 \textrm{N}_4$ integrated photonic network (Quix Quantum BV). One of the fiber couplers is placed on a linear stage (SLC-$2475$, Smaract GmbH) to achieve temporal overlap of the photons inside the chip. The optical chip consists of $12$ input and $12$ output modes and is fully tunable\cite{Taballione2021}. Once the photons have propagated through the chip, they are routed towards the superconducting nanowire single-photon detectors (SNSPD) (Photon Spot). 
A time tagger (Time tagger ultra, Swabian Instruments) is used to read out the single clicks of the detectors and trigger photodiode (TDA $200$, PicoQuant) and their coincidences. The laser's output power is constantly monitored with a calibrated photodiode.

The single-photon source is pumped with $50\,$mW, which results in a photon pair rate of about $210$ kHz. The heralding efficiency, i.e. the probability of detecting the second photon when the first is detected, is typically around $45\%$. The photons are $93\%$ indistinguishable. The detector dark counts are suppressed with the trigger photodiode to an average of $3.1\pm 0.8\,$Hz.

The integrated photonic chip is a $\textrm{Si}_3 \textrm{N}_4$ netwerk by Quix Quantum. The optical chip consists of 12 input and 12 output modes with a Clements-type network, linking all input and output modes with each other\cite{Clements2016}. The matrices are implemented with an average fidelity of $F = \tfrac{1}{n} \textrm{Tr}(U_{\textrm{t}}U^*_{\textrm{e}}) = 0.69 \pm 0.07$. 

The losses in the chip are low (<0.1 dB/cm)\cite{Roeloffzen2018a} and dominated by the fiber-to-chip connections which are around $20\%$ per facet. Furthermore, any losses on the chip are uniformly distributed over the modes because of the square geometry of the chip. This is important as it allows to divide out the optical losses and describe the propagation of the photons through the chip by a unitary matrix\cite{Oszmaniec2016}.

\section{Supplemental Materials}
    \label{sec:supmat}

\paragraph{Simulation of scattering systems}

The simulation of the 12-mode $S$-matrices follows the model as proposed by Dorokhov, Mello, Pereira and Kumar\cite{Dorokhov1982,Mello1988}, which divides the scattering system into short segments. Each segment is shorter than the transport mean free path $l^*$ and longer than the wavelength. Adding a new segment can now be described as a perturbative correction\cite{Rotter2017}. We follow the transfer method of Ko and Inkson\cite{Ko1988} for numerical stability.

In our case, the matrices are computed by simulating a one-dimensional $6$-mode waveguide with perfectly reflecting boundary conditions. The waveguide is divided into $40$ equally sized sections over the length of the waveguide. Each section has a probability of $10\%$ to have a scatterer placed at a random coordinate inside this waveguide segment. This probability corresponds to the weak scattering regime.

The probability to encounter a scatterer relates to the transport mean free path $l^*$. Furthermore, given $l$, the number of segments $N$ determines the average transmission efficiency. In our case, we chose $N=40$ and $\langle T \rangle =0.37$ as this allows us to observe open channels with $200$ random instances of these waveguides. Stronger-scattering waveguides, i.e. with more segments and scatterers, have lower average transmission such that an insufficient number of singular values can be sampled to resolve the open channels. The limit of $200$ matrices is chosen for experimental convenience.

\paragraph{Matrix embedding}
 Our matrix embedding procedure with the no-gain assumption implies that all singular values of the \textit{entire} $S$ matrix should be smaller than or equal to $1$. This is achieved by embedding the $6 \times 6$ transmission matrix inside a larger $12\times 12$ matrix\cite{Shen2017}. In this section we describe this procedure.
 
The $n \times n$ scattering matrix $S$ can always be decomposed by the singular value decomposition: $\textrm{svd}(S) = U \Sigma V^*$. Here, the unitary matrices $U$ and $V^*$ describe some basis transformation to the eigenvectors of the matrix. The matrix $\Sigma$ is a diagonal matrix with the singular values and describes the 'weight' 
of the eigenmode. Ideally, the network is lossless and as a result, $S$ must be unitary. In that case, the diagonal elements of $\Sigma$ are all of the form of $e^{i\theta_n}$, with $\theta_n$ some phase of the $n^{\textrm{th}}$ singular value. In case the amplitude of a singular value is $<1$, there the corresponding eigenmode is lossly and if it is $>1$, then it has gain.

In our case, we only have access to a noisy version of the transmission matrix. The noise eludes the observation of the open channels, so it is essential to mitigate the noise on $T$. We achieve this by embedding the transmission matrix $T$ inside a larger, unitary matrix which is constructed using the matrices of the singular value decomposition. This is necessary as it is not possible to impose unitarity. The process of embedding the transmission matrix inside a larger, unitary matrix can physically be understood by interpreting loss as a beam splitter where one of its output modes directs the light to an unobserved, inaccessible mode\cite{Brod2020}. The exact splitting ratio corresponds directly to the loss in the system. This indicates that the larger unitary matrix should be at least twice the size of the physical system, so that each mode can have at least one loss channel available. Note that gain is nonphysical in our system as there is no additional light source present, besides the injected photons.

The goal now is to first construct a new unitary matrix $S'$ that incorporates the coupling to the environment and then to impose the no-gain assumption. For the first step, we construct new matrices $U'$, $\Sigma'$ and $V'$, which together form $S'= U' \Sigma' V'^{*}$, which incorporates the coupling to the environment. Recall that $\textrm{svd}(T) = U \Sigma V^*$. 

The new matrix $U'$ can be constructed by:
\begin{equation}
    U' = \begin{bmatrix}
        U &0\\
        0 &I \end{bmatrix},
\end{equation}
where the bottom right of $U'$ is filled with an identity matrix for convenience. In principle, any unitary matrix can be used as there is no input or output in any of the unobserved modes anyway. The matrix for $V'$ is constructed similarly. 

The matrix $\Sigma'$ now denotes not just the singular values, but also the coupling to the unobserved modes, i.e., the loss channels. This results in four quadrants, each quadrant is a diagonal matrix. The off-diagonal quadrants denote the coupling to the environment and are constructed such that the energy is conserved in a $L_2$ (Euclidian) norm. The new matrix is given by

\begin{equation}
    \Sigma' = \begin{bmatrix}
    \begin{array}{c | c}
        D_i & D_o\\\hline
        D_o & D_i\\
    \end{array}
    \end{bmatrix},
    \label{eq:eqSigma'}
\end{equation}
with $D_i$ and $D_o$ matrices given by:
\begin{equation}
    D_i = \begin{bmatrix}
        \cos{^2\theta_1}& 0& \ldots& 0 \\
        0& \cos{^2\theta_2}& & \vdots&\\
        \vdots& & \ddots& 0\\
        0& \ldots& 0& \cos{^2\theta_n}\\
    \end{bmatrix},\\
\end{equation}
and
\begin{equation}
    D_o = \begin{bmatrix}
        \sin{^2\theta_1}& 0& \ldots& 0 \\
        0& \sin{^2\theta_2}& & \vdots&\\
        \vdots& & \ddots& 0\\
        0& \ldots& 0& \sin{^2\theta_n}
    \end{bmatrix}.
\end{equation}

In these matrices, $\cos{^2\theta_i} = \tau'$, with $\tau'$ the singular values after the no-gain restriction. The no-gain restriction entails that no $\tau>1$ as it is nonphysical in our system. Hence the maximal allowed singular value is $\tau = 1$, meaning that all singular values of a matrix should be rescaled to $\tau' = \tau/\textrm{max}\tau$. Imposing {\it no gain} in the transmission values is a sufficient error correction strategy for suppressing experimental noise and retrieving the bimodal distribution without going so far as imposing unitarity on S. These renormalised singular values are then used to compute $\Sigma'$ in Eq. \ref{eq:eqSigma'}. The resulting new singular values are shown in Fig. \ref{fig:fig3DataRaw}b) of the main text.

\paragraph*{Acknowledgements}
We acknowledges funding from the Nederlandse Wetenschaps Organisatie (NWO) via QuantERA QUOMPLEX (Grant No. 680.91.037), and Veni (grant No. 15872). Furthermore we would like to thank Allard Mosk for discussions and Klaus J. Boller for proofreading.

\paragraph*{Author contributions statement}

PP conceived the project, RM detailed and performed the experiment and analysed the data, Quix Quantum provided reduced cross talk settings for its chip. All authors reviewed the data and reviewed and contributed to the manuscript. 

\paragraph{Competing interests} Both P.W.H. Pinkse and J.J. Renema are shareholders of Quix Quantum BV.

\bibliography{DMPK_references}

\end{multicols}
\end{document}